\documentclass[twocolumn,pra,aps,superscriptaddress]{revtex4-1}
\usepackage[T1]{fontenc}
\usepackage[latin9]{inputenc}

\usepackage{amsmath}
\usepackage{amssymb}
\usepackage{xcolor}
\usepackage{bbm}
\usepackage{braket}
\usepackage{mathrsfs}  
\usepackage[normalem]{ulem}
\allowdisplaybreaks

\usepackage{graphicx}

\usepackage[colorlinks=true,linkcolor=blue,urlcolor=blue,citecolor=blue]{hyperref}
\begin{document}
\title{Selective branching, quenching, and converting of topological modes}
\author{Toshikaze Kariyado} 
\affiliation{International Center for Materials Nanoarchitectonics, National Institute for Materials Science, Tsukuba, Ibaraki 305-0044, Japan}
\author{Robert-Jan Slager}
\affiliation{TCM Group, Cavendish Laboratory, University of Cambridge, J. J. Thomson Avenue, Cambridge CB3 0HE, United Kingdom}

\begin{abstract}
A salient feature of topological phases are surface states and many of the widely studied physical properties are directly tied to their existence. Although less explored, a variety of topological phases can however similarly be distinguished by their response to localized flux defects, resulting in the binding of modes whose stability can be traced back to that of convectional edge states. The reduced dimensionality of these objects renders the possibility of arranging them in distinct geometries, such as arrays that branch or terminate in the bulk. We show that the prospect of hybridizing the modes in such new kinds of channels poses profound opportunities in a dynamical context. 
In particular, we find that creating junctions of  $\pi$-flux chains or extending them as function of time can induce transistor and stop-and-go effects. Pending controllable initial conditions certain branches of the extended defect array can be actively biased. Discussing these physical effects within a generally applicable framework that relates to a variety of established artificial topological materials, such as mass-spring setups and LC circuits, our results offer an avenue to explore and manipulate new transport effects that are rooted in the topological characterization of the underlying system.

\end{abstract}
\maketitle

\section{Introduction}
With the discovery of time reversal symmetry (TRS) invariant topological insulators (TIs)~\cite{RevModPhys.82.3045,RevModPhys.83.1057}, the study of topological materials has become a prominent subject of interest in condensed matter physics, resulting in a plethora of phases and characterizations~\cite{Clas1a,Clas1b,Clas1c,Clas2,Shiozakicrystal, Hughesmagnetic, Turnermagnetic,Clas3,Clas4,Bouhonglobal,Clas5,unal2020quench,obs2,Tra2019,Weylsemimetal,Bouhon2019,bzduvsek2016nodal,yang2014classification}. A notable signature of many topological phases, and hence one of the main motivations to study them, is the bulk-boundary correspondence~\cite{prodan2016bulk,Hatsugai,khalaf2019boundaryobstructed, Slager2015, Rhim2018, boundarymodes, song2017, benalcazar2017quantized, PhysRevB.99.041301,trifunovic2020higherorder,schindler2018higher}. 
The anomalous edge states, circumventing the Nielsen-Ninimiya theorem~\cite{nielsen}, in many cases directly relate to sought-after physical signatures, ranging from axion behavior \cite{QiAxion} to possible majorana excitations~\cite{Fumajorana}, and accordingly also hold promise for future technological applications~\cite{Moore}.

On another note, many TIs can also be characterized by their response to symmetry preserving gauge fluxes that act as monodomy defects. Such fluxes bind midgap modes that in turn relate to the underlying topological invariant~\cite{Jackiw,PhysRevD.24.2669,PhysRevLett.98.186809,PhysRevB.77.033104,ran2008spin,qi2008spin,PhysRevLett.108.106403,mesaros2013zero,Tra2019,RevModPhys.89.041004,Wu_2014}. These gauge fluxes, which need to be localized to a plaquette of the order of a lattice constant, are generally hard to realize experimentally, although routes that use defects to mimic the effective response~\cite{ran2009one,imura2011weak,PhysRevLett.108.106403,PhysRevLett.123.266802, PhysRevB.93.245406,PhysRevB.90.241403, PhysRevB.93.245406}  are increasingly proving viable in experimental context~\cite{bisbdis,Nayakeaax6996}. In this regard, the controllability of artifical materials \cite{Huber_metamaterials} or cold atom systems~\cite{RevModPhys.83.1523,PhysRevLett.120.243602} could provide for a more immediate route to access this physics. Indeed, in a recent study we considered $\pi$-flux arrays in such meta-material settings~\cite{PhysRevResearch.1.032027} and showed that the hybridization processes can be tuned to induce arbitrary flat semi-metallic bands.

\begin{figure}[h]
 \centering
 \includegraphics[scale=1.0]{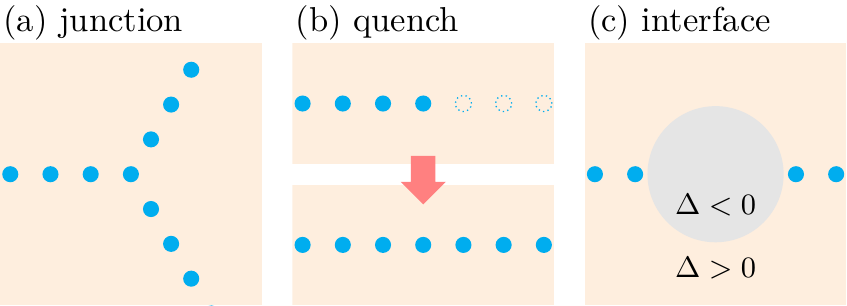}
 \caption{Target setups. (a) $\pi$-flux chain junction. (b) Terminated $\pi$-flux chain and quenching. (c) Interaction between $\pi$-flux chain and topological interface modes.}
 \label{fig:scope}
\end{figure}

In this work we take maximum advantage of both the highly tuneable nature of artificial materials and the flexibility of $\pi$-flux chains to uncover new dynamical effects. In contrast to normal interfaces such chains in the bulk of topological materials offer different possibilities of terminating as well as branching them. Using experimentally implemented models, we in particular consider the target geometries illustrated in Fig.~\ref{fig:scope}. That is, we discuss the dynamics in topological meta-materials across chain junctions [Fig.~\ref{fig:scope}(a)], in terminated chains that are extended upon a quench [Fig.~\ref{fig:scope}(b)] and finally also in geometries that connect the chains to interface modes [Fig.~\ref{fig:scope}(c)]. As a main result we find that, pending the initial conditions, the intensity across one branch can be substantially biased with regard to the other in the different set-ups, rather akin to a classical transistor. In addition, we also find a stop-and-go effect in the transmission upon dynamically extending the $\pi$-flux chain. Given the routinely implementable control of the parameters in the systems we discuss, we believe that our work can set a basis to experimentally explore these new bulk effects in a wider context.

\section{Model and Method}
For concreteness we depart from the experimentally well studied modulated honeycomb model~\cite{Yves:2017aa,Barik666,Yu:2018aa,Li:2018aa,Noh:2018aa,PhysRevLett.120.217401,Zhirihin_2018,Chaunsali_2018,freeney2019edgedependent}, however we emphasize that for each geometry we also present an effective theory, see Appendices B and C, that further underpins the generality and offers additional routes of implementing our results. The according Hamiltonian reads
$H=\sum_{\langle ij\rangle}t_{ij}c^{\dagger}_{i}c_{j}$,
in which $\langle ij\rangle$ indicate nearest neighbours and $c^{\dagger}_i$  refers to the creation operator at a site $i$.
The hopping terms $t_{ij}$ feature a modulation set by $t_0=\bar{t}+\delta$ and $t_1=\bar{t}-2\delta$. The former relates to the black bonds, while the latter to the red bonds in Figs.~\ref{fig:junction}-\ref{fig:interface}(a) and are distributed as follows. Starting from a specific hexagon one assigns black bonds to all its edges connecting the six sites. Subsequently, all bonds that depart from this hexagon are modulated as red bonds. Finally, the next hexagons to which the red bonds connect are again all assigned black bonds as for the first step. This  
results in a structure of isolated hexagons having black bonds connected by red bonds to six other such hexagons. [See the inset of Fig.~\ref{fig:junction}(a).]

The effective Dirac cones of the honeycomb model are then gapped out due to the modulation and the according mass term, which features a sign that is set by that ratio $|t_0|$ and $|t_1|$ for a fixed sign of $\delta$. These insulating phases can in fact also be considered in more detail \cite{PhysRevLett.114.223901,Wu:2016aa,Kariyado:2017aa} and, using mirror winding numbers, be more adequately characterized from a crystalline perspective. 

After fixing the hopping term geometry $t_{ij}$, $\pi$-fluxes may be introduced by flipping the sign of selected $t_{ij}$. This is achieved by placing hypothetical strings on the system, which change the sign of hoppings on the bonds crossing the string, resulting in $\pi$-fluxes at their end points. For the gapped phase with $\delta>0$, when $\pi$-flux threads through a hexagon of black ($t_0$) bonds, it induces two in-gap modes per $\pi$-flux. These in-gap modes are protected by the sublattice (chiral) symmetry and the mirror symmetry whose reflection plane is perpendicular to the zigzag direction as was to be expected given that the bulk topology is captured by mirror winding numbers. In other words, for an isolated $\pi$-flux, the two in-gap modes are indexed by the chiral and the reflection parity indices. Namely, one of the states has its wave function weight exclusively on the A-sublattice [sublattices 1-3 in the inset of Fig.~\ref{fig:junction}(a)] and is reflection even, while the other has its weight exclusively on the B-sublattice [sublattices 4-6 in the inset of Fig.~\ref{fig:junction}(a)] and is reflection odd (or, vice versa)~\cite{PhysRevResearch.1.032027}.

We remark that in some studies of this type of modulated honeycomb lattice model, the $\delta<0$ phase is denoted topological. However, at the level of the effective Dirac theory, this concept is only relative. That is, it only indicates that the mass term changes its sign across the phase boundary between $\delta>0$ and $\delta<0$. Here, we consider $\pi$-fluxes in the $\delta>0$ phase, which resonates better with a reflection-symmetric placement of $\pi$-fluxes than in the $\delta<0$ phase in terms of generating the in-gap modes.

As a next step, we consider the dynamics, which is governed by 
\begin{equation}
    \frac{d^2 x}{dt^2}=-\hat{\Gamma}x + f^{(c)}\cos\Omega t + f^{(s)}\sin\Omega t, \label{eq:master}
\end{equation}
That is, the system is characterized by a dynamical matrix $\hat{\Gamma}$ due to the application of an external monochromatic (single frequency) force.
Here, $x$ is an $N$-dimensional vector for a system with $N$ degrees of freedom.
The matrix elements of $\hat{\Gamma}$ are determined through $\Gamma_{ij} = \frac{\partial^2 V}{\partial x_i\partial x_j}$ with
\begin{equation}
    V = \frac{1}{2}\sum_{i\in S}\sum_{j\in S,E} k_{ij}(x_i-x_j)^2, 
\end{equation}
where $i\in S$ represents the summation over sites. In order to apply a fixed boundary condition, which is important to preserve the sublattice symmetry of the honeycomb lattice at the boundary \cite{Kariyado:2015aa}, the summation also runs over hypothetical external sites $E$ that have $x_j=0$ ($j\in E$). As a result, we then obtain
\begin{equation}
    \Gamma_{ij} = \sum_{l\in S,E} k_{il}\delta_{ij}-k_{ij},
\end{equation}
where we match $k_{ij}$ to the Hamiltonian of the modulated honeycomb lattice tight-binding model, i.e., $k_{ij}=t_{ij}$, such that underlying background is indeed the modulated honeycomb model.

From a physical point of view Eq.~\eqref{eq:master} describes any system that is essentially a coupled harmonic oscillator system. This covers, for instance, spring-mass models or LC-circuits. Accordingly, $x$ then may represent the displacement of mass points, or the voltage at a certain point in a circuit, while $V$ plays the role of elastic energy in  spring systems or charging energy of circuit elements. 

Pursuing this analogy further, the external force is attained by singling out a $\pi$-flux threaded hexagon and applying  $f_0\cos\Omega t$ ($f_0\sin\Omega t$) on the sublattice 1 (sublattice 4) in the unit hexagon. We note that this gives a quarter period phase shift between the A-sublattice localized $\pi$-flux mode and the B-sublattice localized $\pi$-flux mode, or more specifically, the input on the B-sublattices is quarter period phase advanced. Note that the quarter period phase shift corresponds to the phase factor $i$ in quantum cases. 

Using this setup we can then study the dynamics under the applied force in the different $\pi$-flux chain geometries alluded to above. However, before moving to this main part, we first introduce the following quantity that will act as a guiding observable,
\begin{equation}
    I_i = \frac{1}{2}\Omega^2x_i^2+\frac{1}{2}\Bigl(\frac{dx_i}{dt}\Bigr)^2.
\end{equation}
This quantity roughly corresponds to an intensity on each site and hence allows for tracking of the effective dynamics.
For convenience, we accordingly also make use of $I^{\text{hex}}_a$, which is defined as the sum of $I_i$ within a unit hexagon specified by an index $a$. 

\section{Results}
With the definitions in place we now turn the subject of the dynamics in the different $\pi$-flux chain setups of Fig.~\ref{fig:scope}. In the static case, each $\pi$-flux threaded plaquette binds a pair of modes whose stability is rooted in the topological characterization of the underlying model~\cite{PhysRevResearch.1.032027}.
Their origin can be understood as being a consequence of the existence of edge states. Hence, upon applying a force, it is to be anticipated that these degrees of freedom, hybridized over the different chain setups, can be manipulated. 
This opens up routes to new effects in combination with the versatility of arranging single $\pi$-flux threaded as opposed to edges hosting surface states.

In all cases below, we assume that the system is at rest at $t=0$. Also, we always use hexagon 1 to inject energy
with the quarter period phase shift between the A- and B- sublattices as noted above. 
We set the frequency $\Omega$ of the external force as $\Omega=\sqrt{3}+0.01$, where $\sqrt{3}$ is the middle of the bulk gap and 0.01 is added to prevent possible nonessential effects by the mixed right movers and left movers. Finally, we use $\bar{t}=1$ and $|\delta|=0.2$.

\subsection{Junction as transistor}
\begin{figure}
    \centering
 \includegraphics[scale=1.0]{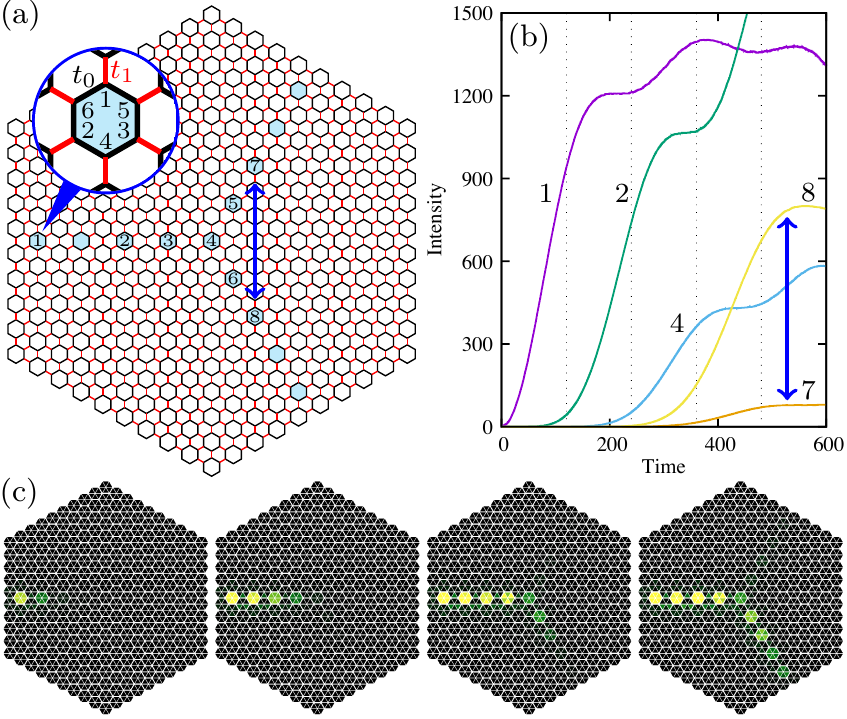}
    \caption{(a) Schematic picture for a system with a $\pi$-flux chain junction. The cyan colored unit hexagons are threaded with $\pi$-fluxes. The system contains 13 $\pi$-fluxes, and eight of them are numbered for convenience. The inset shows the unit cell structure and the sublattice indices 1-6. (b) Time evolution of intensity $I^{\text{hex}}_i$ for $i$th numbered hexagon. The intensities for the 7th and 8th hexagons show contrasting behavior. (c) Snapshots of the intensity map. The four shots correspond to the time slices indicated by the vertical dotted lines in (b). Bright color means high intensity. Majority of the energy from the source flows into the lower branch.}
    \label{fig:junction}
\end{figure}

Turning first to the junction, we find the numerical results displayed in Fig.~\ref{fig:junction} for the full model. The intensity is transmitted from the source (hexagon 1) to the right along the $\pi$-flux chain [Figs.~\ref{fig:junction}(b) and (c)]. Interestingly, after passing the junction, the intensity shows a biased behavior; the lower branch has significantly larger intensity than the upper branch. We have checked that the bias is inverted if the force on the B-sublattice in hexagon 1 is delayed by a quarter period phase rather than advanced, and that there is no bias without the phase shift in the input (Appendix C). That is, the fate of the $\pi$-flux mode is controlled by the local manipulation of the input at the source. We therefore conclude that the junction behaves rather akin a transistor, whose bias can be manipulated in a controlled fashion. 

In order to understand this phenomenon qualitatively, let us consider how the intensity is transmitted from hexagon 3 to hexagon 5 or 6 over the junction. Owing to the chiral symmetry of the model, the only relevant matrix elements are those coupling the A and B sublattices. Therefore, starting from hexagon 3, the intensity can be passed to the A-sublattice sites of hexagon 6 in two ways. Either via the process
$h_{A}^{3}\rightarrow h_{B}^{4}\rightarrow h_{A}^{6}$ or $h_{B}^{3}\rightarrow h_{A}^{6}$, where $h_{\alpha}^{j}$ denotes the $\alpha=A,B$ sublattice of hexagon $j$.
Together with the phase acquired along the real space paths, the inherent quarter period phase shift between A- and B-sublattices at the origin leads to the constructive interference at hexagon 6. Furthermore, if we try to relate hexagon 5 and 6,
the roles of the A- and B-sublattices are interchanged due to the lattice geometry, and the advance (delay) is reinterpreted as delay (advance). This consequently results in destructive interference at hexagon 5, concluding the essential explanation underpinning the bias between the branches. The same course of thinking also elucidates why the bias is inverted or suppressed upon changing the phase shift at the source. This relation furthermore brings to light that the biasing ability is determined by the ratio between the coupling
of $h^3_{A,B}\rightarrow h^4_{B,A}\rightarrow h^{5,6}_{A,B}$ and $h^3_{A,B}\rightarrow h^{5,6}_{B,A}$, which will be affected by the bulk gap.

Crucially, this analysis can be enforced by analyzing the descriptive model named the projected model as detailed in Appendix B. There we construct an effective tight-binding model for $\pi$-flux modes by projecting the system onto the subspace spanned by the in-gap states. This procedure conveys the effective hoppings between the $\pi$-fluxes, which satisfactory reproduces the dynamics of the full model.
Eliminating some of the hopping processes in the projected model by hand, it is then readily confirmed that when the hoppings between the hexagon 3, 5, and 6 are turned off, the biasing behavior is also turned off. This is in line with the above consideration. 

\subsection{Termination and quench}
\begin{figure}
 \centering
 \includegraphics[scale=1.0]{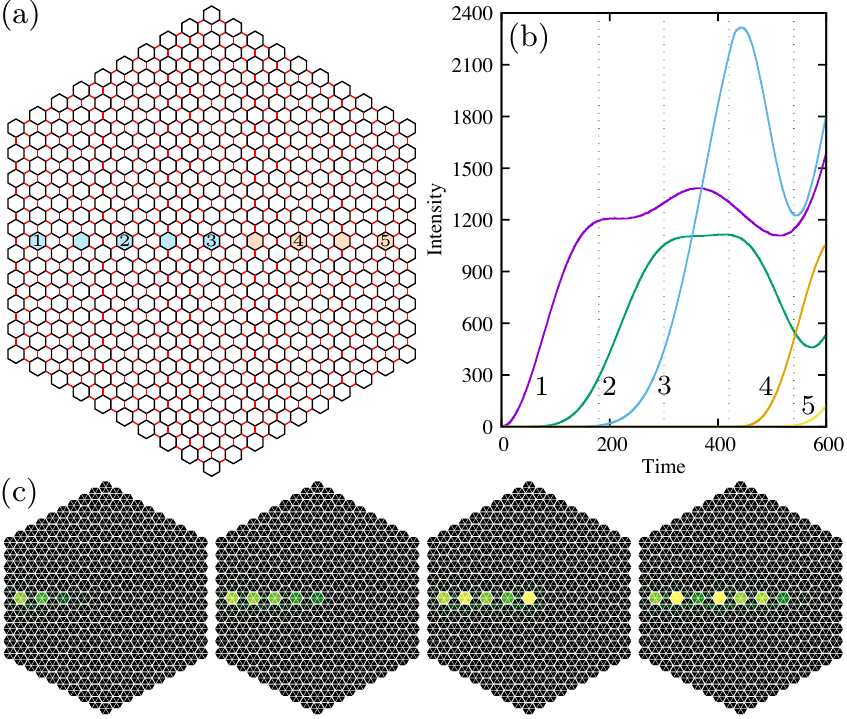}
 \caption{(a) Schematics for the quench setup. The cyan colored hexagons are always threaded by $\pi$-fluxes, while the orange colored hexagons are threaded by $\pi$-fluxes only in the second stage in the time line. The second stage starts at $t=420$, at the time corresponding to the third vertical dotted line in (b). (b) Time evolution of intensity $I^{\text{hex}}_i$ for $i$th numbered hexagon. We can see the delay of intensity rise at the hexagon 4. (c) Snapshots of the intensity map. The four shots correspond to the time slices indicated by the vertical dotted lines in (b). The third one is at the moment that the second stage starts. 
 }
 \label{fig:quench}
\end{figure}

Armed with the junction results, let us move on to the terminated chain that is extended via a quenching procedure, defining two time domains. In the first time domain, $t\in[0,400)$, $\pi$-fluxes thread the cyan colored hexagons only, while in the second time domain, $t>400$, they thread both the cyan and the orange colored hexagons, see~Fig.~\ref{fig:quench}(a). We assume that the change at $t=400$ is instantaneous and use the final state in the first domain as the initial state in the second domain in the numerical evaluation. We find that the intensity evolves as in Figs.~\ref{fig:quench}(b) and \ref{fig:quench}(c). The key observation is that, although the distances are the same, the delays of the intensity growth from hexagon 2 to 3 and 3 to 4 are different. The intensity at hexagon 4 only grows after $t=400$, that is in the second time domain. This thus shows that we can realize a stop-and-go process along the $\pi$-flux chain. 

It is interesting to interpret this result in terms of the band structure along the $\pi$-flux chain. In the previous study \cite{PhysRevResearch.1.032027}, we have shown that the band structure for the in-gap $\pi$-flux modes along the $\pi$-flux chain has two 1D Dirac cones, one at the Brillouin zone center and the other at the zone edge. Because of this, the $\pi$-flux modes can be back-scattered on the chain even between the same pseudospin component. This is in sharp contrast with the topological interface states that cannot be back-scattered without (pseudo)spin flips. Similarly, interfaces do not admit a termination at the middle of the bulk. We thus conclude, at the cost of being back-scattered, the $\pi$-flux modes enables new device building blocks culminating in controlled branching or stop-and-go mechanisms. 

\begin{figure}
 \centering
 \includegraphics[scale=1.0]{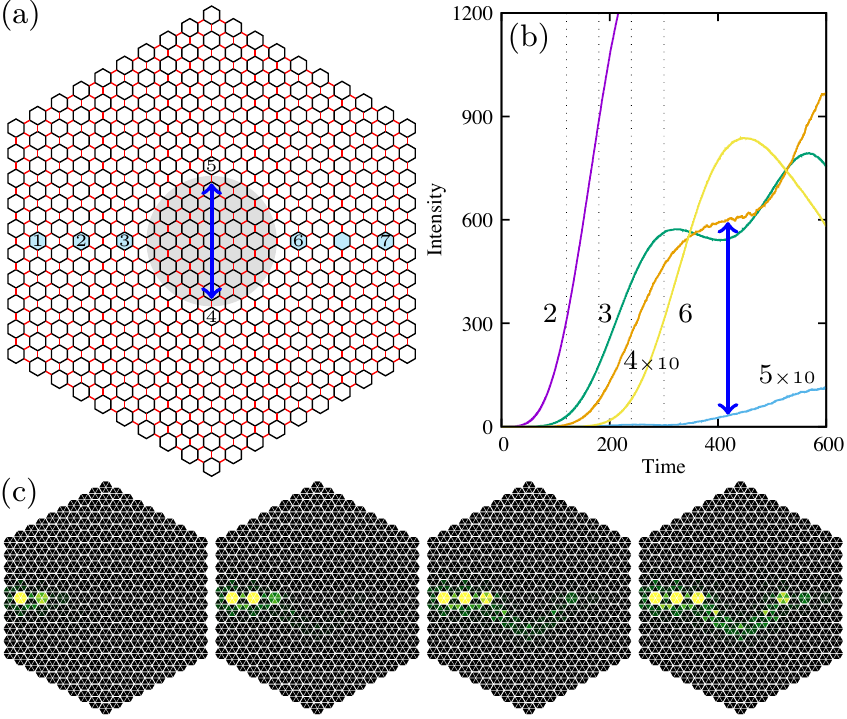}
 \caption{(a) System with $\pi$-flux chains and a topological interface. In the circular shaded region at the center, we set $\delta=-0.2$, while in the other region, we set $\delta=0.2$. (b) Time evolution of intensity $I^{\text{hex}}_i$ for $i$th numbered hexagon. For better comparison, $I^{\text{hex}}_4$ and $I^{\text{hex}}_5$ are scaled by a factor of 10. The intensities for the 4th and 5th hexagons show contrasting behavior. (c) Snapshots of the intensity map. The four shots correspond to the time slices indicated by the vertical dotted lines in (b). Majority of the energy from the source flows along the lower half of the interface.}
 \label{fig:interface}
\end{figure}

\subsection{Coupling to interface modes}
Finally, we consider situations involving the $\pi$-flux chain and interface states, as in the setup of Fig.~\ref{fig:interface}(a). The parameter $\delta$ is positive outside the gray shaded region as before, while $\delta$ is negative inside of the gray shaded region. The $\pi$-fluxes are again threading the cyan colored hexagons. The resulting time evolution of the intensity is summarized in Figs.~\ref{fig:interface}(b) and \ref{fig:interface}(c). One thing we notice is that the intensity propagates faster along the interface than along the $\pi$-flux chain, indicated by the timing of the intensity grow at hexagon 3 and 4. The difference in the group velocity along the interface and along the $\pi$-flux chain will generally depend on the bulk gap size. The other, more striking, thing we notice is that there is a significant difference between the intensity at hexagon 4 and hexagon 5. In Fig.~\ref{fig:interface}(c), we can see that the intensity propagates along the lower half of the circular interface. As in the case of biasing the $\pi$-flux chain junction, the way that intensity branches is linked to the phase shift at the source. If delay and advance is exchanged, the intensity goes around the upper half of the interface, or if there is no phase shift, the intensity splits equally to the upper and the lower halves of the interface, see Appendix C. We conclude that, as in the junction, we can bias branching channels, now involving edge states degrees of freedom, by manipulating the initial condition at the source offering new switch effects in the bulk of the topological material. 
In order to understand this branching, it is essential to know the overlap between the wave functions for the $\pi$-flux modes and the interface modes. For the $\pi$-flux modes, the wave functions can be inferred from the results in the isolated $\pi$-flux case \cite{PhysRevResearch.1.032027}. For the interface modes, a rough description of the wave functions is obtained by approximating the bulk states by a $k\cdot p$ type continuous model (see Appendix D). It turns out that the phase delay or advance in the $\pi$-flux modes effectively selects the right or left movers in the interface states to couple.

\section{Conclusion and discussion}
In conclusion, we have shown that the versatility of $\pi$-fluxes in topological meta-materials leads to new dynamical effects. In particular, we find that branched or extendable channels can be designed, which can be biased pending on the initial conditions. Most profoundly, this culminates in transistor and stop-an-go effects that are  rooted in the topology of the underlying bulk.

These mechanisms maximally profit from the engineerability, positively impacting the implementabilty of the above results in three manners. Firstly, the modulated honeycomb system can be readily realized in disk-spring systems or LC-circuits. Secondly, these systems offer realistic handling of the initial conditions being the real control parameter to establish and manipulate the branching effects. Thirdly, they offer a rather feasible route to create $\pi$-flux threaded plaquettes.  Indeed, 
Ref.~\cite{Kumar2019} for instance pointed out how one could control the hopping sign essential for implementing such fluxes using reversed springs. Similarly, specific kinds of wiring as proposed in Ref.~\cite{Schuster2019} induce a route to implement them in LC circuits. Finally, we point out that yet another route is to control the potential of neighbors, meaning that nearest hopping terms can be replaced by an intermediate site \cite{PhysRevResearch.1.032027}. Adjusting the relative strength of the potential can then induce a sign flip with respect to the initial hopping process.

In experiments, long-distance propagation along the $\pi$-flux chain requires small dissipation. Such propagation along the interface has presently been confirmed in experiments. Comparing with the interface modes, the group velocity is smaller for the modes in the $\pi$-flux chain with the current parameter choice, as we have noted. Sharing the frequency with the interface modes, the smaller group velocity thus requires a higher Q-factor for the $\pi$-flux chain to have the same propagation distance. However, the group velocity can be controlled by the bulk gap size; smaller bulk gap leads to larger real space extension of each $\pi$-flux mode, resulting in larger interflux hoppings and a larger group velocity. That is, at the cost of being more restrictive in the frequency space, we can relax the Q-factor requirement. Practically, the parameter can then be optimized to balance these factors.

In view of the experimental implementablity and  general nature of our results, which apply to a wide variety of systems, we believe that our results could propose an interesting avenue to explore in the near future.

\begin{acknowledgments}
 R.-J.~S. acknowledges funding via the Marie Sk{\l}odowska-Curie programm under EC Grant No. 842901 and the Winton programme as well as Trinity College at the University of Cambridge.
This work was furthermore supported by JSPS KAKENHI Grant Number JP17K14358 (T.K.). The part of computation in this work has been done using the facilities of the Supercomputer Center, the Institute for Solid State Physics, the University of Tokyo. 
\end{acknowledgments}
\bibliography{dynamics}

\clearpage
\newpage
\appendix
\section{Solution by normal mode decomposition}
We briefly comment on Eq.~\eqref{eq:master}. Introducing a basis transformation $\hat{O}$ that diagonalizes $\hat{\Gamma}$, this Equation reduces to 
\begin{equation}
   \frac{d^2\tilde{x}_l(t)}{dt^2}=-\omega_l^2\tilde{x}_l(t)+\tilde{f}^{(c)}_l\cos\Omega t + \tilde{f}^{(s)}_l\sin\Omega t, 
\end{equation}
where $\tilde{\bm{x}}=\hat{O}^\dagger\bm{x}$, $\tilde{\bm{f}}^{(c,s)}=\hat{O}^\dagger\bm{f}^{(c,s)}$ and $\omega_l^2$ are the eigenvalues of $\hat{\Gamma}$. This system can be solved upon introducing an ansatz 
\begin{equation}
 \tilde{x}_l(t) = a_l^{(c)}\cos\Omega t+a_l^{(s)}\sin\Omega t + b_l^{(c)}\cos\omega_lt+b_l^{(s)}\sin\omega_lt.
\end{equation}
Given the initial state $\{\tilde{x}_l(t_{\text{init}}),\frac{d\tilde{x}_l(t_{\text{init}})}{dt}\}=\{x_l^{(0)},v_l^{(0)}\}$, the four parameters $a^{(c)}$, $a^{(s)}$, $b_l^{(c)}$, and $b_l^{(s)}$ are fixed by the four inputs $x_l^{(0)}$, $v_l^{(0)}$, $\tilde{f}^{(c)}_{l}$, and $\tilde{f}^{(s)}_{l}$. 

\section{Projection to the subspace spanned by the $\pi$-flux modes}
We here detail the effective tight-binding model to underpin the results of the main text. We stress the generality of the presented effective theory that as such offers a broad perspective to implement physics in a variety of systems. We first outline the effective description of the $\pi$-flux modes and then turn these results into representative description for the systems considered in the main text.

\subsection{Projecting method}
Our first goal is to construct an orthonormalized basis set $[w]=\{|w_i\rangle\}$ that (i) spans the subspace spanned by an orthonormalized basis set $[\psi]=\{|\psi_l\rangle\}$ ($l\in\text{target}$), and (ii) inherits the structure of a set of $N=\mathrm{dim}[\psi]=\mathrm{dim}[w]$ candidate functions, $[w^{(c)}]=\{|w^{(c)}_i\rangle\}$, in the best possible manner. We require $[w^{(c)}]$ to be normalized, but not necessaliry orthogonalized. $[w^{(c)}]$ may \textit{not} span $[\psi]$ completely. Typically, $[\psi]$ is a set of selected eigenfunctions of a given Hamiltonian, while $[w^{(c)}]$ is a set of localized functions selected by some intuition or physical consideration. 

The transformation between $[w]$ and $[\psi]$ has to be unitary, i.e., we should have
\begin{equation}
 |w_i\rangle=\sum_{l\in\text{target}}(X)_{li}|\psi_l\rangle, \quad X^\dagger X=\hat{1}, 
\end{equation}
to preserve the orthonormalized condition. 
For the following use, an overlap matrix $q$ is defined as
\begin{equation}
 (q)_{li} = \langle \psi_{l}|w^{(c)}_i\rangle.
\end{equation}
Applying singular value decomposition (SVD), we have $q = U\Lambda V^\dagger$
where $U$ and $V$ are unitary and $\Lambda$ is diagonal with nonnegative diagonal elements. 

Now, we fix $X$ so as to make $\{|w_i\rangle\}$ and $\{|w^{(c)}_i\rangle\}$ as similar as possible. The similarity is measured by 
\begin{align}
 \Delta &= \sum_i (\langle w_i|-\langle w^{(c)}_i|)(|w_i\rangle-|w^{(c)}_i\rangle),\\
 &= 2N-2\mathrm{Re}\sum_{i}\langle w_i|w^{(c)}_i\rangle,
\end{align}
where the second line follows from the normalization of $[w^{(c)}]$. 
Then, smaller $\Delta$, or equivalently larger $\mathrm{Re}\Xi$ with $\Xi=\sum_{i}\langle w_i|w^{(c)}_i\rangle$, means better similarity between $[w]$ and $[w^{(c)}]$. Now, $\Xi$ is evaluated as
\begin{equation}
 \Xi
  = \sum_{i,l} (X)_{l,i}^*(q)_{li}
 = \mathrm{Tr}(\Lambda M)=\sum_i\lambda_i(M)_{ii} 
\end{equation}
where $\lambda_i$s are diagonal elements of $\Lambda$ and $M=V^\dagger X^\dagger U$. By definition, $M$ is unitary, which implies $|(M)_{ii}|\leq 1$, since $\sum_j|(M)_{ij}|^2=1$. Furthermore, since $\lambda_i$s are nonnegative real numbers, maximum $\textrm{Re}\Xi$ is achieved when $(M)_{ii}=1$, which forces $(M)_{ij}=0$ ($i\neq j$). From the above, we should have $M=\hat{1}$ for maximizing $\Xi$. Note that $M=\hat{1}$ means $X=UV^\dagger$, and the optimized $\{|w_i\rangle\}$ becomes
\begin{equation}
 |w_i\rangle = \sum_{l\in\text{target}}(UV^\dagger)_{li}|\psi_l\rangle.
\end{equation}

\subsection{Dynamics in the projected model}
To derive an effective tight-binding description for the states in the $\pi$-flux chain, we use the numerically obtained eigenstates for the set of in-gap states as $[\psi]$, while we use the analytically obtained wave functions for the $\pi$-flux modes at $|t_1/t_0|\rightarrow 0$ limit as $[w^{(c)}]$ \cite{PhysRevResearch.1.032027}. With this choice, each of the wave functions in $[\psi]$ is not necessarily well localized because it is eigenstate and there are effective hoppings between $\pi$-fluxes as we derive, while each of the wave functions in $[w^{(c)}]$ is well localized since there is no coupling between the $\pi$-fluxes in the $|t_1/t_0|\rightarrow 0$ limit. Consequently, the hoppings in the effective model are essentially short ranged, which is useful in intuitive understanding. For convenience, we name this effective tight-binding model the {\it projected model}.

\begin{figure}
 \centering
 \includegraphics[scale=1.0]{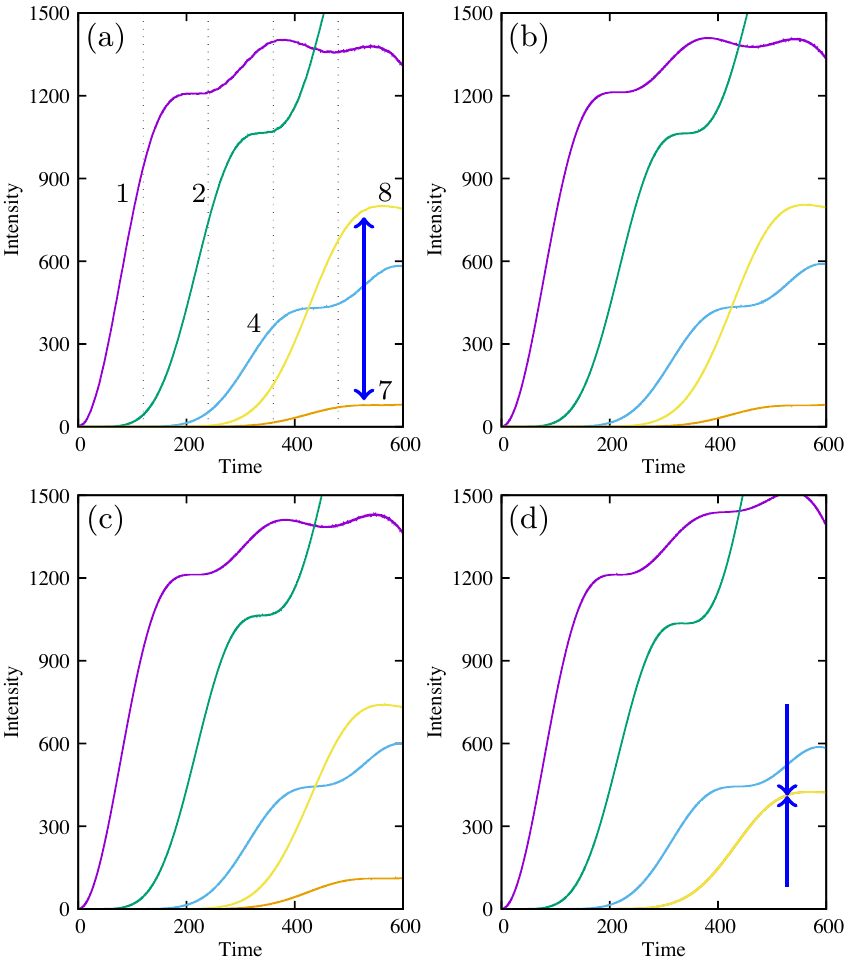}
 \caption{(a) Time evolution of $I^{\text{hex}}_i$ for the full model, reproduced from Fig.~\ref{fig:junction}(b). See the caption of Fig.~\ref{fig:junction} for notations. (b) Time evolution of $I^{\text{hex}}_i$ for the projected model, showing good agreement with the full model. The minor difference is from the coupling to the bulk modes eliminated in the projected model. (c) The same as (b), but preserving hoppings in the projected model between only upto the next nearest neighbor pairs of $\pi$-flux threaded hexagons, and truncating the others. Still showing good agreement with the full model. (d) The same as (c), but only preserving hoppings between the nearest neighbor pairs of $\pi$-flux threaded hexagons. Imbalance between the hexagons number 7 and 8 collapses.}
 \label{fig:comparison}
\end{figure}

Now, let us see how the projected model performs. Fig.~\ref{fig:comparison}(a) displays the results for the full model, which is thus reproducing Fig.~\ref{fig:junction}(b) for comparison. In Fig.~\ref{fig:comparison}(b) we then show the results of the projected model, matching the full model closely. In order to extract the essence of the biasing phenomenon, we truncate minor hoppings in the projected model. Even if we preserve hoppings only between the nearest neighbor pairs of the $\pi$-fluxes (like hexagon 3 and 4) and the next nearest pairs of the $\pi$-fluxes (like hexagon 3 and 5, 6), the model maintains the satisfactory agreement with the full model [See Fig.~\ref{fig:comparison}(c)]. However, if we further switch off the hoppings between the next nearest pairs of the $\pi$-fluxes, the biasing goes away [See Fig.~\ref{fig:comparison}(d)]. This supports the theory described in the main body. 

\section{Effect of phase shift}
\begin{figure}
 \centering
 \includegraphics[scale=1.0]{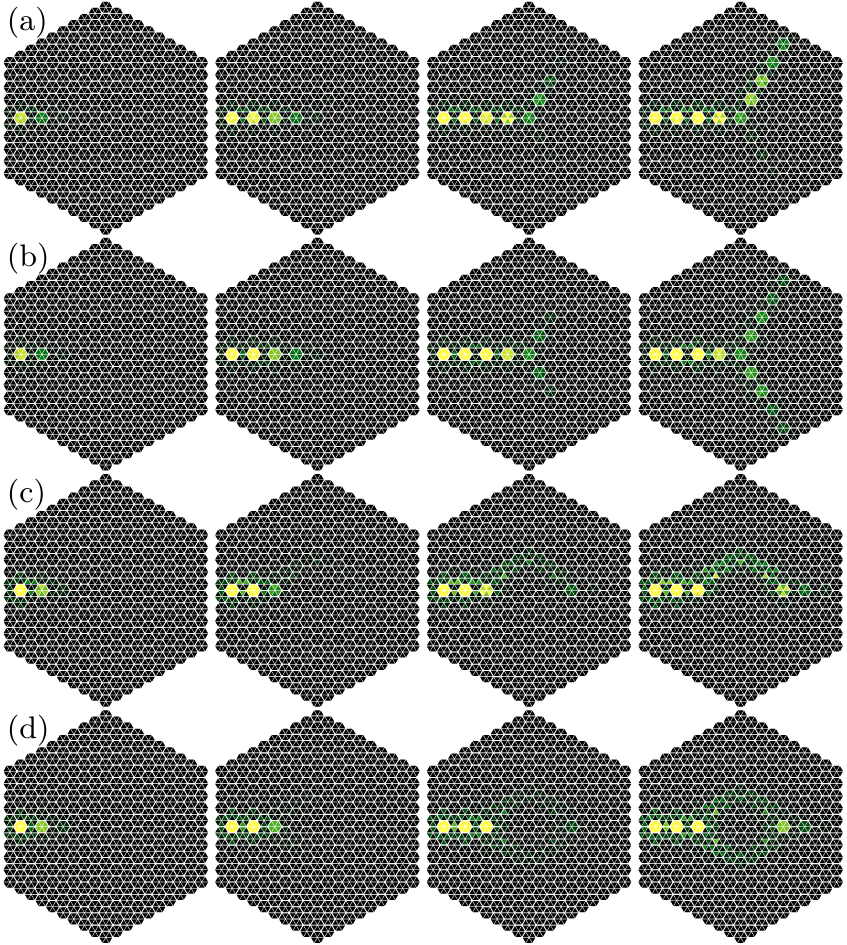}
 \caption{(a) Snapshots for the $\pi$-chain junction with the opposite quarter period phase shift in the input. Majority of the energy flows into the upper branch, instead of the lower branch in Fig.~\ref{fig:junction}. (b) The same as (a), but without phase shift in the input. The energy splits into the both branch equally. (c) Snapshot for the $\pi$-flux chain and the topological interface with the opposite quarter period shift in the input. Majority of the energy flows along the upper half of the interface. (d) The same as (c), but without phase shift in the input. The energy splits into the both upper and lower half of the interface.}
 \label{fig:phase_shift}
\end{figure}

Here we check the effects of the phase shift in the input for the junction and the coupling to the interface. Figure~\ref{fig:phase_shift}(a) is for the junction with delay and advance being exchanged from the main body. We see that the intensity mainly goes to the upper branch after the junction. Figure~\ref{fig:phase_shift}(b) is also for the junction but without phase shift. There is no biasing in this case. Figures~\ref{fig:phase_shift}(c) and \ref{fig:phase_shift}(d) are for the coupling to the interface, with delay and advance exchanged and without phase shift. As for the junction, the way that intensity flows is changed by the delay/advance switching, and there is no bias without phase shift. 

\section{Coupling to the interface states}
The coupling between the $\pi$-flux states and the interface states is determined by the overlap between the wave functions of these states. Here we recall the very similar, and hence intricately related, form of the edge states and the $\pi$-flux modes as detailed in Refs.~\cite{PhysRevLett.108.106403,PhysRevResearch.1.032027}. In order to derive the wave function for the interface state, we use a continuum $k\cdot p$ model around the $\Gamma$-point. Following the notations in Ref.~\cite{PhysRevResearch.1.032027} and its Appendix, and using the indexing of the sublattices in the inset of Fig.~\ref{fig:junction}(a), the according Hamiltonian becomes
\begin{equation}
    H_k=
    \begin{pmatrix}
    H_+(k)&0\\
    0&H_-(k)
    \end{pmatrix},
\end{equation}
where $H_{\pm}(k) = \Delta\sigma_z\pm vk_x\sigma_x+vk_y\sigma_y$ 
for the basis set $\{|\psi_1\rangle,|\psi_2\rangle,\mathcal{K}|\psi_1\rangle,\mathcal{K}|\psi_2\rangle\}$ with
\begin{equation}
    |\psi_1\rangle=\frac{1}{\sqrt{6}}
    \begin{pmatrix}
    1\\
    \omega^*\\
    \omega\\
    -1\\
    -\omega^*\\
    -\omega
    \end{pmatrix},\,
    |\psi_2\rangle=\frac{1}{\sqrt{6}}
    \begin{pmatrix}
    1\\
    \omega\\
    \omega^*\\
    1\\
    \omega\\
    \omega^*
    \end{pmatrix}.
\end{equation}
Here, $k_\pm=k_x\pm ik_y$, $\omega=-\frac{1}{2}+\frac{\sqrt{3}}{2}i$, and $\mathcal{K}$ denotes the complex conjugate operator. 
Furthermore, the mass term $\Delta$ is of the form $C+v^2(k_x^2+k_y^2)$, where $C$ is a constant, see also \cite{PhysRevResearch.1.032027}.

\begin{figure}
    \centering
    \includegraphics{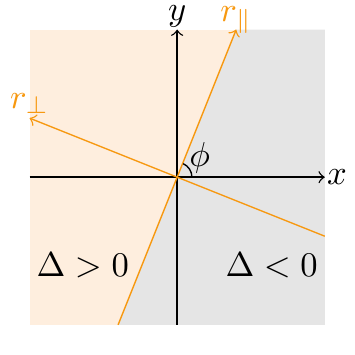}
    \caption{Schematic picture of the interface making angle $\phi$ with the $x$-axis.}
    \label{fig:angle_theta_interface}
\end{figure}
For the set up described in Fig.~\ref{fig:angle_theta_interface}, by transforming the basis set to $\{|\alpha_\phi\rangle,|\beta_\phi\rangle,\mathcal{K}|\alpha_\phi\rangle,\mathcal{K}|\beta_\phi\rangle\}$ with
\begin{align}
    |\alpha_\phi\rangle &= \frac{1}{\sqrt{2}}\bigl(
    e^{-\frac{i\phi}{2}}|\psi_1\rangle+e^{\frac{i\phi}{2}}|\psi_2\rangle
    \bigr),\\
    |\beta_\phi\rangle &= \frac{1}{\sqrt{2}}\bigl(
    e^{-\frac{i\phi}{2}}|\psi_1\rangle-e^{\frac{i\phi}{2}}|\psi_2\rangle
    \bigr),
\end{align}
and replacing $\Delta$ and $k_\perp$ as $\Delta\rightarrow \Delta_0 \mathrm{sgn}(r_\perp)$ and $k_\perp\rightarrow -i\partial_{\perp}$, the Hamiltonian becomes $\tilde{H}(k_\parallel,r_\perp)=\tilde{H}_+(k_\parallel,r_\perp)\oplus \tilde{H}_-(k_\parallel,r_\perp)$ with
\begin{equation}
    \tilde{H}_{\pm}(k_\parallel,r_\perp)=
    \begin{pmatrix}
    \pm vk_\parallel&\Delta_0\mathrm{sgn}(r_\perp)-v\partial_\perp\\
    \Delta\mathrm{sgn}(r_\perp)+v\partial_\perp& \mp vk_\parallel
    \end{pmatrix},
\end{equation}
to the first order in $|k|$. When $\Delta_0/v > 0$, a physical solution for $\tilde{H}_\pm$ is
\begin{equation}
    |\psi_{\text{phys}}\rangle \propto 
    \begin{pmatrix}
    e^{-\frac{\Delta_0}{v}|r_\perp|}\\
    0
    \end{pmatrix},\quad
    E=\pm vk_\parallel.
\end{equation}
Namely, the wave function for the right (left) mover is proportional to $|\alpha_\phi\rangle$ ($\mathcal{K}|\alpha_\phi\rangle$).

In Fig.~\ref{fig:interface}(a), the $\pi$-flux chain hits the armchair section ($\phi=\pi/2$) of the interface. The associated wave function represents itself in the original basis as
\begin{equation}
    |\alpha_{\frac{\pi}{2}}\rangle = \frac{1}{\sqrt{2}}
    \bigl(e^{-\frac{i\pi}{4}}|\psi_1\rangle+e^{\frac{i\pi}{4}}|\psi_2\rangle\bigr)=
    \frac{1}{2\sqrt{6}}
    \begin{pmatrix}
    2\\
    -(1+\sqrt{3})\\
    -(1-\sqrt{3})\\
    2i\\
    -i(1-\sqrt{3})\\
    -i(1+\sqrt{3})
    \end{pmatrix}.
\end{equation}
In this basis, the first three components belong to the sublattice A, while the other components belong to the sublattice B [see Fig.~\ref{fig:junction}(a)]. Then, $|\alpha_{\frac{\pi}{2}}\rangle$ precisely shows the quarter period phase shift between the sublattice A and B. It also indicates that the right and the left movers switch their roles by $\mathcal{K}$, which switches the phase delay and advance.

\end{document}